# Mechanical Properties of Protomene: A Molecular Dynamics Investigation

Eliezer F. Oliveira[1,2], Pedro A. S. Autreto[3], Cristiano F. Woellner[4], and Douglas S. Galvao[1,2]

[1]Gleb Wataghin Institute of Physics, University of Campinas - UNICAMP, Campinas, SP, Brazil.
[2]Center for Computational Engineering & Sciences (CCES), University of Campinas - UNICAMP, Campinas, SP, Brazil.
[3]Center of Natural Human Science, Federal University of ABC - UFABC, Santo Andre, SP, Brazil.
[4]Department of Physics, Federal University of Paraná - UFPR, Curitiba, PR, Brazil.

ABSTRACT

*Recently, a new class of carbon allotrope called* protomene *was proposed. This new structure is composed of $sp^2$ and $sp^3$ carbon-bonds. Topologically, protomene can be considered as an $sp^3$ carbon structure (~80% of this bond type) doped by $sp^2$ carbons. First-principles simulations have shown that protomene presents an electronic bandgap of ~3.4 eV. However, up to now, its mechanical properties have not been investigated. In this work, we have investigated protomene mechanical behavior under tensile strain through fully atomistic reactive molecular dynamics simulations using the ReaxFF force field, as available in the LAMMPS code. At room temperature, our results show that the protomene is very stable and the obtained ultimate strength and ultimate stress indicates an anisotropic behavior. The highest ultimate strength was obtained for the x-direction, with a value of ~110 GPa. As for the ultimate strain, the highest one was for the z-direction (~25% of strain) before protomene mechanical fracture.*

## INTRODUCTION

The mixture of different carbon hybridized states can generate several new materials with different properties and dimensionalities [1], such as fullerenes (0D), carbon nanotubes (1D), and graphene (2D). Some of these new carbon-based structures are built mixing $sp^2$ and $sp^3$ carbon states in the same material. Following this ideas, L. A. Burchfield and co-workers [2] hypothesized a new carbon-based material called novamene. This material is based on the combination of the hexagonal diamond (carbon $sp^3$ bonds) and hexagonal carbon rings (carbon $sp^2$ bonds). From theoretical calculations [2,3], this new structure presents an indirect bandgap of ~0.3 eV, and its ultimate strength can reach values ~ 100 GPa, which make novamene an interesting material for electromechanical applications.

Using similar ideas, F. Delodovici and co-workers [4] proposed another carbon allotrope based on the mixture of $sp^2$ and $sp^3$ hybridized states, which they called protomene. Protomene is also based on the hexagonal diamond structure doped by $sp^2$ bonds, but in a fewer amount when compared to novamene. From first principles calculations, protomene presents a direct bandgap of ~3.38 eV, comparable to the observed for gallium nitrides (GaN) [4]. This result suggests that protomene could be a good candidate to replace GaN in some electronic applications. However, its mechanical properties, which can be of fundamental importance to some of these applications, have not been yet investigated. These properties are one of the objectives of the present work. We carried out a detailed study of protomene mechanical properties under tensile strain through fully atomistic molecular dynamics simulations.

**MATERIAL AND METHODS**

All molecular dynamics (MD) simulations were carried out with ReaxFF force field [5], as implemented in the computational code LAMMPS [6]. In Figure 1 we present the unit cell of protomene: a hexagonal crystalline arrangement with $a_0 = b_0 =$ 8.07 Å and $c_0 = 3.51$ Å, lattice angles of $\alpha=\beta=90°$ and $\gamma=120°$, and 48 carbon atoms. The carbon atoms colored in red (darker) and green (lighter) represent the atoms in $sp^2$ and $sp^3$ hybridizations, respectively.

For the MD simulations, we used a protomene hexagonal supercell with a size of $5a_0$ x $5b_0$ x $5c_0$ (6000 atoms) with applied periodic boundary conditions (PBC). In order to avoid any residual structural stress before the tensile calculations, the protomene geometry was initially energetically minimized using a conjugate gradient technique, then followed by a thermal equilibration at 300 K in an NPT ensemble at a constant pressure of 0 GPa. In this thermal equilibration, the protomene structure showed to be stable, with no bond-type interconversions and/or structural failures. In order to obtain some of the protomene mechanical properties (ultimate strength, ultimate stress, and Young's modulus), after the thermal equilibration, we applied a uniaxial tensile strain, at a rate of $10^{-6}$ $fs^{-1}$ along the Cartesian x, y, and z-directions (see Figure 1) to deform the protomene structure.

For each direction of the tensile process, we use the corresponding normal stress to produce the stress-strain curves. During the externally applied uniaxial tensile strain, we kept the system at 300 K and external pressure of 0 GPa for the tensile perpendicular and shear directions using an NPT ensemble. We also analyzed the stress distribution for each atom during the tensile process using the standard von Mises stress analyses, which are very helpful to understand the fracture mechanisms in protomene. For all calculations, we used a timestep of 0.25 fs.

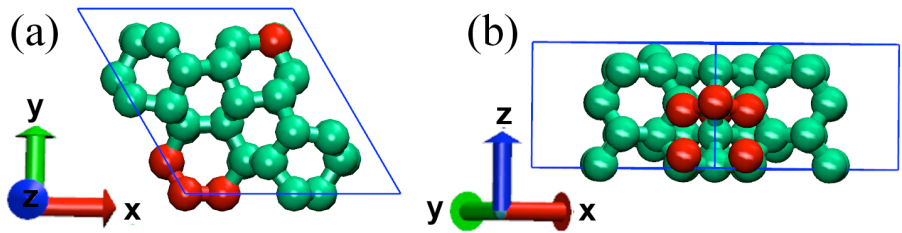

Figure 1. Unit cell of protomene: (a) top and (b) side of view. The atoms highlighted in red (darker) and green (lighter) are bonded in $sp^2$ and $sp^3$ hybridizations, respectively.

## RESULTS AND DISCUSSIONS

In Figure 2 we present the stress-strain curves for protomene when stretched along the x, y and z directions at 300 K. We can see from this Figure that the curves suggest that protomene has an anisotropic behavior regarding the tensile mechanical response. A large elastic regime is present in all curves, followed by an abrupt drop in stress values, which is suggestive of a brittle behavior. For a strain of about 10%, the stress-strain curves for x and y directions are similar, but with different ultimate strength values (110.1 and 96.7 GPa for x and y directions, respectively). As for z-direction, smaller ultimate strength is observed (86.8 GPa). However, the ultimate strain for z-direction is higher (reaching 24.7%) in comparison to the corresponding x and y ones (17.7% and 18.8%, respectively). From the linear regime of the stress-strain curves, it is also possible to estimate the protomene Young's modulus. The obtained values were 658.2, 665.6, and 475.5 GPa for x, y, and z directions, respectively. These different values are consistent with the general anisotropic behavior inferred from Figure 2.

For comparisons, we present in Table 1 the ultimate strength and strain, and Young's modulus of protomene, graphene (armchair), and carbon nanotube (12,0) at 300 K. The mechanical properties of these others carbon allotropes were also obtained using the ReaxFF force field [7]. As can be seen, without considering the ultimate strain for protomene z-direction, the protomene values are in general smaller than these others carbon allotropes. The Young's modulus of protomene suggests a more deformable material than the graphene and nanotube. However, its values for ultimate strength and strain are still higher than some other materials, such as some ceramics, silicon or titanium alloys [8]. F. Delodovici and co-workers [4] suggested that protomene could be a good candidate to replace GaN in electronic applications. Experimentally it is found an ultimate strength and strain of around 10 GPa and 18%, respectively for GaN [9,10], lower than the same observed for protomene. These values suggest that protomene could be a more resilient material than GaN.

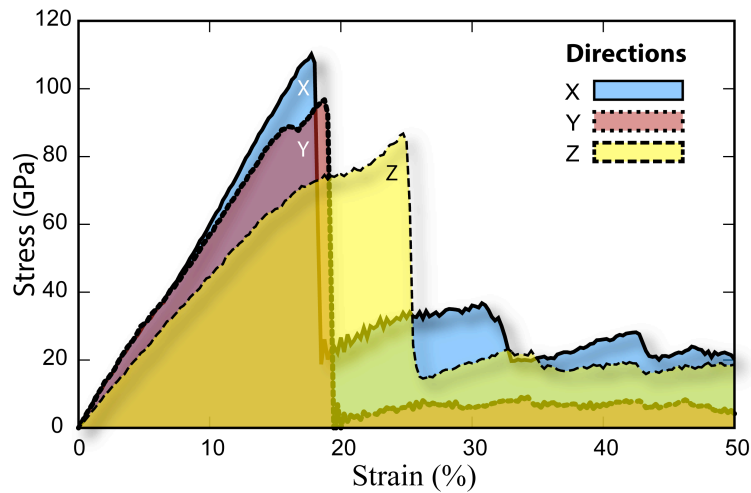

Figure 2. Stress-strain curves for protomene when a uniaxial stress applied in the x, y, and z directions.

Table 1: Mechanical properties for protomene (this work) and some selected carbon allotropes [7] obtained with a ReaxFF force field.

| Structure | Ultimate Strength (GPa) | Ultimate Strain (%) | Young's Modulus (GPa) |
|---|---|---|---|
| Protomene (x-direction) | 110.1 | 17.7 | 658.2 |
| Protomene (y-direction) | 96.7 | 18.8 | 665.6 |
| Protomene (z-direction) | 86.8 | 24.7 | 475.5 |
| Graphene (armchair) | 149.6 | 21.5 | 1266.0 |
| Nanotube (12,0) | 111.3 | 21.6 | 1214.0 |

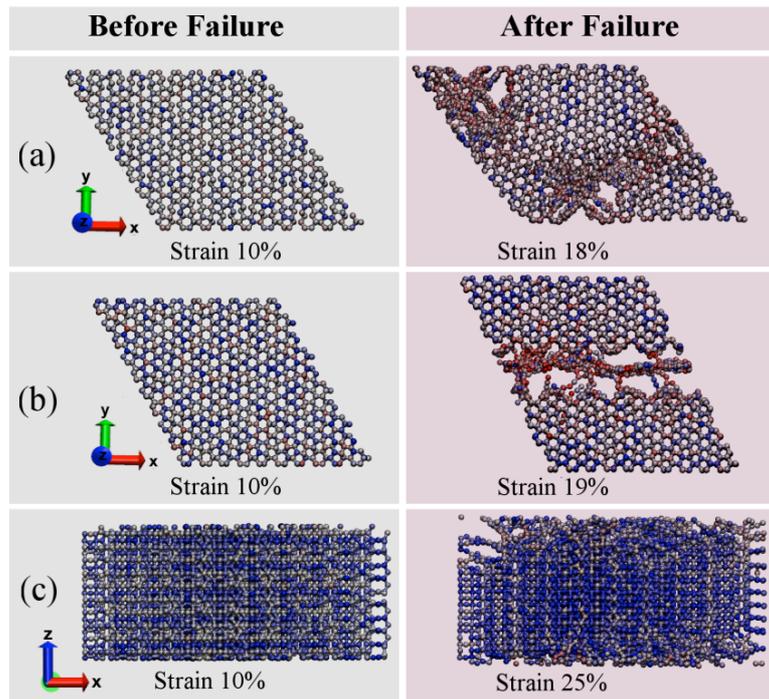

Figure 3. MD snapshots (before and after structural failure) of stress distribution in protomene for uniaxial tensile procedure along (a) x, (b) y, and (c), z directions. The atoms are colored accordingly to its von Mises stress values (vary from blue/lighter (low) to red/darker (high)).

In Figure 3 we present MD snapshots of the protomene at different strain levels during the tensile procedure along x-, y-, and z-direction and at 300 K. The atoms are colored accordingly to its von Mises stress values (vary from blue/lighter (low) to red/darker (high)). We can see that during the tensile deformations the stress is well-distributed in protomene, independently of the tensile direction. When the strain reaches values close to the protomene mechanical failure (fracture), the stress becomes more concentrated on the carbon atoms connected by sp$^3$ bonds. Along x- and y-directions, the sp$^3$ bonds are located mostly in different planes, which are more likely to break during the tensile process. For the z-direction, the more stressed sp$^3$ atoms are those that are parallel to the tensile direction. Our MD results suggest that the fracture/crack initiation and propagation originate from the regions in which a high amount of sp$^3$ bonds break at the same time that with the continuously stretching, leads to the material completely structural failure.

## CONCLUSIONS

Our MD results suggest that protomene has an anisotropic behavior about tensile deformation. At room temperature, we observed that the protomene ultimate strength (~100 GPa) and Young's modulus (~600 GPa) are lower than other carbon allotropes (such as graphene and nanotubes). However, with relation to the protomene z-direction deformation it presents a higher ultimate strain (~24.7%), suggesting that it can support larger deformations along this direction. Although these values are smaller than the corresponding ones for other carbon allotropes (such as graphene and carbon nanotubes), they still outperform other materials, such as some ceramics, silicon, steel, GaN, or titanium alloys.

With relation to the mechanical failure (fracture), the protomene is also anisotropic. For x- and y-directions, the sp$^3$ bonds are located mostly in different planes, which are more likely to break during the tensile process, while that for the z-direction, the more stressed sp$^3$ atoms are those that are parallel to the tensile direction [11]. A similar analysis can be also considered as if the xy plane was composed of graphene-like structures, while along the z-direction just diamond-like ones.

## ACKNOWLEDGMENTS

The authors acknowledge support from the Brazilian agencies CNPq, CAPES, and FAPESP (Grant 2016/18499-0). The authors also thank the Center for Computational Engineering and Sciences at UNICAMP for financial support through the FAPESP/CEPID Grant 2013/08293-7.

**References**

1. T. D. Burchell, *Carbon materials for advanced technologies*, 1st ed. (Elsevier Science, Oxford, 1999).
2. L. A. Burchfield, M. A. Fahim, R. S. Wittman, F. Delodovici, and N. Manini, *Heliyon* **3**, e00242 (2017).
3. E. F. Oliveira, P. A. S. Autreto, C. F. Woellner, and D. S. Galvao, *Carbon* **139**, 782 (2018).
4. F. Delodovici, N. Manini, R. S. Wittman, D. S. Choi, M. A. Fahim, and L. A. Burchfield, *Carbon* **126**, 547 (2018).
5. A. C. T. van Duin, S. Dasgupta, F. Lorant, and W. A. Goddard, *J. Phys. Chem. A* **105**, 9396 (2001).
6. S. J. Plimpton, *Comput. Phys.* **117**, 1 (1995).
7. B. D. Jensen, K. E. Wise, and G. M. Odegard, *J. Comput. Chem* **36**, 1587 (2015).
8. R. E. Smallman and A. H. W. Ngan, *Physical metallurgy and advanced materials engineering*, 7td ef. (Elsevier, Butterworth-Heinemann, 2007).
9. Q. Peng, C. Liang, W. Ji, and S. De, *Appl. Phys. A* **113**, 483 (2013).
10. T. H. Sung, J. C. Huang, J. H. Hsu, and S. R. Jiang, *Appl. Phys. Lett.* **97**, 171904 (2010).
11. E. F. Oliveira, P. A. S. Autreto, C. F. Woellner, and D. S. Galvao, to be published.